\documentstyle[aps,epsfig,preprint]{revtex}
\newcommand{\beq}{\begin{equation}}
\newcommand{\eeq}{\end{equation}}
\newcommand{\beqa}{\begin{eqnarray}}
\newcommand{\eeqa}{\end{eqnarray}}

\begin{document}
\title {
{\small
\hspace{11cm}  LPT Orsay 99/19\\
\hspace{11cm} hep-ph/9903338 }\\ 
\vspace{1cm}
{\bf Event--by--event fluctuations in heavy--ion 
collisions and the quark--gluon
string model}}

\author{
A. Capella, 
$^{1}$,
E. G. Ferreiro
$^{1}$,
A. B. Kaidalov
$^{2}$,}
\address{
$^{1}$
Laboratoire de Physique Theorique,\\
% et Hautes Energies,\\ 
Unit\'e Mixte de Recherche (CNRS) UMR 8627,\\
Universite de Paris XI, Bat.210, 91405 Orsay Cedex, France,\\
$^{2}$
Institute of Theoretical and Experimental Physics, \\
B. Cheremushinskaya 25, 117 259 Moscow, Russia}

\maketitle

\begin{abstract}

 We apply dynamical string models of heavy--ions collisions at high energies
 to the  
 analysis of event--by--event fluctuations. Main attention is devoted to a
 new variable proposed to study "equilibration" in heavy--ions collisions.
 Recent results of the NA49 collaboration 
 at CERN SPS are compared with predictions
 of the Quark--Gluon String Model (QGSM), which gives a good description of
 different aspects of multiparticle production for collisions of nucleons and
 nuclei. It is shown that the new observable and other results of the 
NA49 analysis
 of event--by--event fluctuations are correctly reproduced in the
 model. We discuss dynamical effects responsible
for these fluctuations and give
 the predictions for p--p, p--Pb and 
Pb--Pb collisions at RHIC and higher
 energies.
 
 \end{abstract}
 
\section{Introduction}

  An event--by--event analysis of heavy--ions collisions can give important
information on the 
dynamics of these processes. In the paper \cite{Gazd1} it was
proposed to use an 
event--by--event analysis of transverse momentum fluctuations as a method
 for the
study of "equilibration" in high--energy nucleus--nucleus collisions.
For this purpose a special variable $\Phi$
has been introduced in ref. \cite{Gazd1} (the definition of this variable will
be given below). The problem of "equilibration" in
high--energy heavy--ions collisions is very important in order to
 understand the
conditions for quark--gluon plasma formation. Recent experimental results on
event--by--event analysis of Pb--Pb collisions at CERN SPS by the
NA49 collaboration
 \cite{NA49} show that the value of $\Phi$ is substantially smaller than 
  expected in the case of independent nucleon--nucleon collisions and
 its smallness was considered as an indication of "equilibration" in the
 system. 
 
 This result has been discussed in the 
framework of different theoretical models.
 It was shown \cite{Gazd2} that the increase of transverse momenta 
 of hadrons due to multiple rescatterings leads to a substantial increase of
 $\Phi$.
Incorporating this effect in the model of ref. \cite{Leon}
 one finds an even stronger disagreement with the NA49 result. 
It was also demonstrated
% \cite{Liu} 
that strings fusion
also leads to an increase of $\Phi$ in disagreement with experiment 
 \cite{Liu,Ele}. 
The results on the influence of final states interactions on the observable
$\Phi$ are contradictory: in ref. \cite{Liu} it was shown that final states 
interactions in the framework of the string model of ref. \cite{Liu}
have a small 
effect and do not allow to reach agreement
with experiment, while in ref. \cite{Bleicher} it was argued that final states
interactions in the framework of the UrQMD model are essential and can
decrease $\Phi$ to a
 value consistent with experiment. On the
other hand, it was shown in ref. \cite{Mrow} 
that, in the case of fully equilibrated
hadronic gas made mostly of pions, one expects large positive values for
the variable $\Phi$ not consistent with the experimental observation.

In this note we study event--by--event fluctuations using the
Monte--Carlo formulation \cite{Amelin} of the Quark--Gluon String Model (QGSM)
 \cite{Kaid1} . The 
QGSM and the Dual Parton Model (DPM) \cite{Capella}) are closely related
dynamical
models based on 1/N--expansion in QCD, string fragmentation
 and reggeon calculus. They 
give a good description of many characteristics of multiparticle production
in hadron--hadron, hadron--nucleus and nucleus--nucleus collisions (for a review
see refs. \cite{Kaid2}). Nuclear interactions in this model are treated
 in the
 Glauber--Gribov approach. It will be shown that the model reproduces the 
results
 of the event--by--event analysis of the NA49 experiment for the quantity
 $\Phi$ as well as for
other fluctuations observed in this experiment \cite{NA49}.
 We analyze the reason for the
 decrease of the quantity $\Phi$ from p--p to Pb--Pb
 collisions seen by the NA49 experiment and come to the conclusion that the
 quantity $\Phi$ is sensitive to many details of the
interaction and can hardly be
 considered as a good measure of "equilibration" in the system. We predict a
 strong increase of the value of $\Phi$ at energies of RHIC and higher. 
The model
 also gives definite predictions for event--by--event fluctuations in
 p--Pb collisions.

\section{Analysis of event--by--event fluctuations of transverse momenta}

Let us remind briefly the method to study
 event--by--event fluctuations of transverse
momenta of produced particles introduced in ref. \cite{Gazd1}. It was proposed
to define for each particle in a given event a variable $z_i=p_{Ti} -<p_{T}>$,
where $p_{Ti}$ is the transverse momentum of the particle $i$ and $<p_{T}>$ is
the mean transverse momentum of particles averaged over all events. Using $z_i$
the quantity $Z=\Sigma^{N}_{i=1} z_i$ is defined, where N is the total number of particles
in the event. 
If nucleus--nucleus collisions can be considered as a superposition
of 
independent nucleon--nucleon collisions then it can be shown \cite{Gazd1} that

\beq
\frac{<Z^2>_{AA}}{<N>_{AA}}=\frac{<Z^2>_{NN}}{<N>_{NN}} \ \ .
\label{eq1}
\eeq
A derivation of this result is given in Appendix 1.

The 
averaging in eq.(1) is over all events in a given kinematical region.
It was proposed in ref. \cite{Gazd1} to characterize the
 degree of fluctuations by
the variable

\beq
\Phi=\sqrt{\frac{<Z^2>}{<N>}} - \sqrt{<z^2>}  \ \ ,
\label{Phi}
\eeq
where $<z^2>$ is the second moment of the single particle inclusive
z--distribution. 
The quantity $<z^2>$
corresponds to purely statistical fluctuations and is determined by mixing
particles from different events. It was emphasized in ref. \cite{Gazd1} that,
 if
nucleus--nucleus collisions would be a simple superposition of independent
nucleon--nucleon collisions,
 then the variable $\Phi$ would be the same as in
the 
nucleon--nucleon case 
(see Appendix 1). In nucleon--nucleon collisions the quantity $\Phi$ is
different from zero due to dynamical correlations and, in particular, 
due to
the dependence of $<p_T>$ on the
number of produced particles. It was proposed
 to attribute a possible decrease of the quantity $\Phi$ in A--A
collisions to
the effects of "equilibration".

Let us note 
that the model of independent N--N
collisions for nucleus--nucleus
interactions 
is an extremely 
oversimplified one. The Glauber model at high energies is
not equivalent to independent N--N
collisions even for N--A
interactions.
 The space--time 
picture of hadron--nucleus interactions at high energies is
 absolutely 
different from a simple picture of successive reinteractions of an
initial 
hadron with nucleons of the
nucleus (see e.g. \cite{Gribov,Kaid3,Capn}). For
nucleus--nucleus interactions there are extra 
correlations \cite{Kaid3,Andreev}.
The model of independent N--N
collisions does not 
even satisfy energy--momentum
conservation,
 as a nucleon of one nucleus can not interact inelastically several
times with nucleons of another nucleus having the same energy at each
interaction.

In the QGSM as well as in DPM, 
the effects of multiple interactions in hadron--nucleus and nucleus
--nucleus collisions are taken into account in the approach based on the 
topological expansion in QCD \cite{Kaid2}. Probabilities of rescatterings
are calculated in the
framework of the Glauber--Gribov theory and multiparticle configurations in
the final state are determined using AGK \cite{Abr} cutting rules.
In these models
 the Pomeron is related to the cylinder type diagrams, which
correspond to the production of 
two chains of particles due to decays of two $qq-q$ strings.
Multi--Pomeron exchanges
are related to multi--cylinder diagrams which
which produce extra chains of type $q-\overline{q}$. They
 are especially important in
interactions with nuclei. Fragmentation of strings into hadrons is 
described
according to "regge counting rules" \cite{frag}, which give correct 
triple--regge and double--regge limits of inclusive cross sections. 
Note, however, that in the Monte Carlo version used in this paper,
the Artru--Menessier string fragmentation scheme is implemented instead. 
All conservation 
laws (including energy--momentum conservations) are satisfied in this 
approach.

Let us emphasize that at energies $\sqrt{s}\sim 10$ 
GeV the cylinder--type
diagrams give the dominant contributions for N--N
collisions. Extra 
$q-\overline{q}$ chains due
to multi--cylinder diagrams have rather small length in rapidity (short chains)
and do not lead to substantial 
contributions to particle production.
In nucleon--nucleus 
and nucleus--nucleus collisions, the number of short chains
is strongly increased compared to the nucleon--nucleon case (it is proportional
to a number of collisions) and they should be taken into account in any realistic
calculations of multiparticle production on nuclei. This means that for 
p--A and
A--B
collisions there are extra "clusters" of particles 
(short chains, of type $q-\overline{q}$)
 compared to the nucleon--nucleon interactions "clusters" (long
chains connecting valence quarks and diquarks of the 
colliding nucleons).
 
 So we come to the conclusion that in the relativistic Glauber--Gribov dynamics
the characteristics of final particles in N--A and A--B
collisions can not be
expressed in terms of N--N
collisions only, as it was assumed in ref.\cite{Gazd1},
and eq.(1) is not valid in general. In Appendix 1 we give as an illustrative
example the results in a model with two types of clusters.
This model 
is a generalization of the single cluster model of ref. \cite{Gazd1}, and is
much closer to QGSM and DPM.
 
The results of the Monte Carlo calculation for the quantity $\Phi$ are shown in
Table 1 for p--p, p--Pb and central Pb--Pb collisions at SPS energies 
($\sqrt{s}= 19.4$ 
GeV) and at RHIC ($\sqrt{s}=200$ GeV). Predictions of the
model for $\Phi$ are quite different for these two
energies. At SPS there is
 a strong reduction of the quantity
$\Phi$ for nuclear collisions compared to N--N
collisions, while at RHIC
 the quantity $\Phi$ is predicted to be much larger than at SPS and about
 the same for Pb--Pb and 
p--p collisions. At LHC energies the value of $\Phi$,
obtained in our model, is 160 MeV for 
p--p and even larger for
Pb--Pb 
collisions 
(the Monte Carlo code we use does not allow to calculate
precise values of the correlations at LHC energies due to 
a too large number of particle
produced in each event;
so in Table 1 we give predictions of the model at 
$\sqrt{s} =540$ GeV and $\sqrt{s} =1$ TeV
to show the energy dependence of fluctuations). 
The results for SPS energies
 are in a reasonable agreement with
experimental data of the 
NA49 Collaboration \cite{NA49}. The result for
p--p 
interactions at this energy is even higher than the 
estimate, based on
the
dependence of $<p_{T}>$ on the
number of charged particles, given in
ref.\cite{Gazd1}. The model reproduces this correlation reasonably 
well and
shows that this is not the only source of fluctuations leading to 
a non--zero
value of $\Phi$. We find that the quantity $\Phi$ is sensitive also to 
other types of correlations and, in particular, to the
correlations related to conservation of $p_T$ in the process. 

Let us note that the quantity $\Phi$ at SPS energies is very small and is
defined in eq.(2) as a difference of two large numbers (see Table 1),
 so it is very sensitive to all details of dynamical models. 
Because of its smallness it 
is  difficult to obtain a good accuracy in a Monte Carlo calculation of
 this quantity  (especially for nucleus--nucleus collisions, where the maximum
 statistics possible in the Monte--Carlo is $\approx$ 5000 events). In order to
 increase the statistics and to reduce this uncertainty we give in
Table 1 results obtained for the total rapidity interval, while 
the experimental
data of the NA49 Collaboration were obtained in a fixed rapidity interval
 $4 < y_{\pi}^{lab.} < 5.5$.
The error in the values of $\Phi$ in Table 1 is about 1 MeV
for the lowest energies, and increases at high energy.
Other properties of event by event fluctuations observed by NA49 Collaboration
\cite{NA49} were calculated under the condition of the experiment and 
are  reproduced by the model reasonably well as it is shown in
Fig.1.

It follows from Table 1 that at SPS energies there is an increase of
 the quantity $\sqrt{\frac{<Z^2>}{<N>}}$ from 
p--p to 
Pb--Pb collisions, but
 there is an even larger  
 increase for $\sqrt{<z^2>}$ due to the
increase of $<p_T>$ and
to a change 
in the form of the
$p_T$ distribution. The effect of the
correlations
 between $<p_T>$ and the
number of charged particles due to rescatterings
 is, to a large extent,
 compensated at these energies by energy--momentum
 conservation effects. As a result, we find no dependence of $<p_T>$
 on $n_{ch}$ for 
Pb--Pb collisions at SPS.
 For RHIC energies  a strong increase in the values of both
  $\sqrt{\frac{<Z^2>}{<N>}}$ and especially of
$\Phi$ is predicted (see Table 1).
 At these energies 
the
 increase of average transverse momentum with the
number of rescatterings
becomes very important in p--p
interactions and is reproduced by the
QGSM (Fig. 2a). It is shown in Fig. 2b that an increase of $<p_T>$
with multiplicity is predicted at these energies even for 
Pb--Pb collisions, although the effect
is less pronounced
for
heavy-ions collisions than for p--p. 
At LHC
energies all these effects will be stronger than at RHIC
and will produce an increase of
the quantity $\Phi$ in p--p, p--A and A--A 
collisions as energy increases.

The predictions of the model can be easily tested in future experiments
at RHIC and LHC.

\section{Conclusions}

We have shown, in a Monte Carlo version of the QGSM, that at SPS energies the 
quantity $\Phi$, caracterizing event--by--event transverse momentum 
fluctuations, decreases from $\Phi \sim 9$ MeV in p--p collisions to 
$\Phi \sim 2$ MeV in central Pb--Pb collisions. This result for 
Pb--Pb collisions agrees with the measurement of the NA49 Collaboration. 
In ref. \cite{Gazd1}, such a decrease between p--p and Pb--Pb was considered
to be a test of equilibration of the dense system produced in central heavy ion
collisions. We have obtained the same result
in the framework of an independent string model.

At RHIC energies, we predict an increase in the value of $\Phi$ ($\Phi =
75 \div 80$ MeV). In this case $\Phi$ will be approximately the same 
in p--p and central Pb--Pb collisions. At higher energies the value 
of $\Phi$ is predicted to be larger and to increase from p--p to central
Pb--Pb collisions.

Our analysis 
indicates that the quantity $\Phi$ 
can 
hardly
be considered as a good measure of "equilibration" in the
system. However, it can be used as a sensitive test of
dynamical
models.
\\
 
{\bf Acknowledgments}

This work was
supported in part by INTAS grant 93-0079ext, NATO grant OUTR.LG
971390, RFBR grants 96-02-191184a, 96-15-96740 and
98-02-17463. One of the author (E. G. F.) thanks Fundaci\'on 
Ram\'on Areces of Spain for finantial support.
\\

\section*{Appendix 1}

Here we will consider a simplified model of multiparticle production with two
types of "clusters", which is a generalization of the model of ref.\cite{Gazd1},
where clusters of only a single type were produced. As discussed above, 
these "clusters" 
correspond
to $qq-q$ chains (clusters of the first type) and 
$q-\overline{q}$
chains (clusters of the second type). Nucleon--nucleon collision at
SPS--energies can be described with a good accuracy 
by the production of two clusters
of the first type, while for proton--nucleus and nucleus--nucleus
interactions,
production of the second type of 
clusters is important even in this energy range.
%In nucleus-nucleus 
%collisions situation is more complicated as in this case
%it is necessary 
%to take into account sea--sea chains (clusters of the third
%type), so our model with two types of clusters is appropriate for p--A
%collisions only. 
 For independent production of $k_1$  clusters of the first type and $k_2$ -
of the second type,
 with average transverse momenta $<p_T>_i$ and multiplicity
$<n>_i$ for the cluster $i$,
 the following results for $<Z>$ and $<Z^2>$ can be
obtained:
\def\theequation{A.{1}}
\beq
<Z>_{k_1k_2}=k_1<Z>_1 +k_2<Z>_2
\label{A1}
\eeq
where $<Z>_i=<n>_i (<p_T>_i-<P_T>)$ and 
$<P_T>=\frac{(<k_1><n_1><p_T>_1 + <k_2><n>_2<p_T>_2)}{(<k_1><n>_1+<k_2><n>_2)}$ .

\def\theequation{A.{2}}
$$<Z^2>_{k_1k_2}=k_1<Z^2>_1 +k_2<Z^2>_2 + k_1(k_1-1)<Z>_1^2+$$
\beq
+ k_2(k_2-1)<Z>_2^2 +
 2 k_1 k_2 <Z>_1<Z>_2
 \label{A2}
 \eeq

 The expression for $<Z^2>_{k_1k_2}$ can be rewritten as

\def\theequation{A.{3}} 
 \beq
 <Z^2>_{k_1k_2}
= k_1(<Z^2>_1-<Z>_1^2)+ k_2(<Z^2>_2-<Z>_2^2)+<Z>_{k_1k_2}^2 \ \ .
 \label{A3}
 \eeq

For the 
quantity $\Phi$ in eq.(\ref{Phi}) 
it is important that contrary to the case of a single cluster, 
the
expression 
for $<Z^2>$ contains negative terms proportional to $<Z>_i^2$.

Next 
we take
the average over the number of produced clusters with
some distribution $P_{k_1k_2}$

\def\theequation{A.{4}}
$$<<Z^2>_{k_1k_2}>=\sum_{k_1,k_2} P_{k_1k_2}<Z^2>_{k_1k_2}=$$
\beq
=<k_1>(<Z^2>_1-<Z>_1^2)+
 <k_2>(<Z^2>_2-<Z>_2^2) + <<Z>_{k_1k_2}^2> \ \ .
\label{A4}
 \eeq
  
 In the following we will denote this averaging  simply by $<Z^2>$.
We shall concentrate on p--A collisions. In this case, it is easy to show that 
 the last term in eq.(\ref{A4}) is small and can be 
 neglected. To prove this we note that for p-A collisions,
 $k_2=k_1-2$ \cite{Kaid2}
with $k_1=\overline{\nu}+1$, where $\overline{\nu}$
is the average number of collisions   and, thus,
\def\theequation{A.{5}}
 \beq
 <<Z>_{k_1k_2}^2> - <<Z>_{k_1k_2}>^2= (<k^2_1>-<k_1>^2)(<Z>_1 + <Z>_2)^2 \ \ .
 \label{A8}
 \eeq
 Taking into account that (for fixed impact parameter) the distribution in $k_1$
is of a Poisson type with $(<k^2_1>-<k_1>^2)=c_1<k_1>$ and that\\ 
$ <<Z>_{k_1k_2}>=<k_1><Z>_1+<k_2><Z>_2=<k_1><Z>_1+(<k_1>-2)<Z>_2=0$ we obtain:
\def\theequation{A.{6}}
\beq
<<Z>_{k_1k_2}^2>=\frac{4c_1<Z>_2^2 }{<k_1>} \ \ .
\eeq
For large values of $<k_1>$ this quantity is much smaller than the other terms
in the right--hand side of eq.(\ref{A4}). 
 
   The expressions for the
quantities $<N>$ and $<z^2>$, that enter into the
definition of $\Phi$ are selfevident

\def\theequation{A.{7}}
\beq
<N>=<k_1><n_1>+<k_2><n_2>
\label{A5}
\eeq
 and
\def\theequation{A.{8}}
 \beq
 <z^2>=\frac{<k_1><n>_1<z^2>_1+<k_2><n>_2<z^2>_2}{<k_1><n>_1+<k_2><n>_2} \ \ .
 \label{A6}
 \eeq

 Let us denote $\frac{<Z^2>_i}{<n>_i}$ as $<z^2>_i(1+\delta_i)$ and
 $\frac{<Z>_i^2}{<Z^2>_i}\equiv\gamma_i$, $\frac{<k_2><n>_2}{<k_1><n>_1}\equiv
 \alpha$. Taking into account that $ \delta_i\approx
 2\frac{\Phi_i}{\sqrt{z_i^2}}$ and $\gamma_i$ are much smaller than unity we
 obtain the following approximate expression for $\Phi$:

\def\theequation{A.{9}} 
\beq
\Phi=\frac{[(\delta_1-\gamma_1)<z^2>_1 + (\delta_2-\gamma_2)\alpha <z^2>_2]}
{2\sqrt{A}}
\label{A7}
\eeq
where $ A=(1+\alpha)(<z^2>_1+\alpha <z^2>_2) $.
 
 It is important that the terms proportional to $\gamma_i$ give negative
 contributions to $\Phi$ and can substantially decrease the value of $\Phi$.

 For the case of clusters of the same type ($\gamma_i=0, <Z^2>_1=
<Z^2>_2,  
 \delta_1=\delta_2$) we obtain: 

\def\theequation{A.{10}}
\beq
\Phi=\frac{\delta_1}{2} \sqrt{<z^2>}
\label{A9}
\eeq
both for p--p and p--A. We recover in this way 
the result of ref.\cite{Gazd1}. 
%The model with
% two types of clusters considered above is still a simplified one and can not
% be directly compared to experimental data as even qq--q chains in p--A
% collisions are somewhat different from those for 
%p--p collisions: some of
% these chains connect diquarks of nucleons from nuclei not to valence quarks
% (as for p--p case) but to sea--quarks of the colliding proton and length in
% rapidity of these chains differs from those for p--p collisions.
The discussion in this Appendix has been restricted to 
p--A interactions. 
The situation is more complicated 
in A--B collisions. Actually, even for 
p--A, we do not claim that the effect 
discussed in this Appendix is the main reason for the decrease of $\Phi$, 
obtained in the Monte Carlo calculations (see Table 1),
between p--p and p--A collisions at SPS energies. 
Nevertheless, our example illustrates the important effect that a 
modification of the model (i. e., going from one to two 
types of 
clusters) can have on the quantity $\Phi$.
 \\
 
% \section*{References}

\newpage

\section*{Table captions}

\vskip 0.5cm

\noindent{\bf Table 1.}
The results of the Monte-Carlo calculation for the quantities
$\sqrt{\frac{<Z^2>}{<N>}}$ (MeV), $\sqrt{<z^2>} $ (MeV) and
$\Phi$ (MeV) for
p-p, p-Pb and central Pb-Pb-collisions at SPS
($\sqrt{s}= 19.4$ GeV),
RHIC ($\sqrt{s}=200$ GeV)
and higher 
($\sqrt{s}=540$ GeV and 1 TeV) energies. The results at 1 TeV have 
only an indicative value (see main text).

\newpage

\section*{Figure captions}

\vskip 0.5cm

\noindent{\bf Figure 1.} Event spectra characterising the multiplicity,
transverse momentum and rapidity distribution of
charged particles per event for Pb-Pb collisions at
$P_{lab}=158$ AGeV/c and $b \leq 3.5$ fm in the rapidity interval
$4 \le y \le 5.5$.
The full lines are the Monte Carlo results. Experimental data are from 
ref. \cite{NA49}.
%The lower right figure corresponds to the event spectra for
%the transverse momentum distribution of charged particles in the total
%rapidity interval.

\vskip 0.5cm

\noindent{\bf Figure 2.} The dependence of the average transverse momentum
on 
the
multiplicity of charged particles in the window 
$|\eta| \leq 2.5$ at $\sqrt{s}=200$ GeV
for ${\rm p}-\overline{\rm p}$ collisions
compared to experimental data \cite{UA1}
(Fig. 2a) and
for Pb-Pb central
%($b \leq 3.5$ fm)
collisions
(Fig. 2b).

%\vskip 0.5cm
%
%\noindent{\bf Figure 3.} The dependence of the average transverse momentum
%on multiplicity of charged particles for p-p collisions at $\sqrt{s}=540$ AGeV
%and $|\eta| \leq 2.5$
%compared to experimental data

\newpage

%\begin{center}
%{\bf Table 1}
%\vskip 1cm
%\begin{tabular}{ccccc} \hline
%& $\sqrt{s}$ (GeV) & 
%$\sqrt{\frac{<Z^2>}{<N>}}$ (MeV) & $\sqrt{<z^2>}$ (MeV)  & $\Phi$ (MeV) 
%\\ \hline
%p-p & 19.4 & 244.66 & 235.53 & 9.03 \\ \hline
%p-Pb & 19.4 & 243.45 & 243.00 & 0.45 \\ \hline
%Pb-Pb & 19.4 & 265.55 & 263.19 & 2.36 \\ \hline
%\hline
%p-p & 200 & 387.04 & 310.64 & 76.40 \\ \hline
%p-Pb & 200 & 433.76 & 367.83 & 65.93 \\ \hline
%Pb-Pb & 200 & 508.92 & 429.42 & 79.50 \\ \hline
%\hline
%p-p & 540 & 450.11 & 323.57 & 126.54 \\ \hline
%p-Pb & 540 & 524.29 & 397.73 & 126.56 \\ \hline
%Pb-Pb & 540 & 622.57 & 475.18 & 147.36 \\ \hline
%\hline
%p-p & 1000 & 455.5 & 324.4 & 131  \\ \hline
%p-Pb & 1000 & 524.5 & 397.5 & 127 \\ \hline
%Pb-Pb & 1000 & 704.2 & 484.3 & 220 \\ \hline
%\hline 
%\end{tabular}
%\end{center}

\begin{center}
{\bf Table 1}
\vskip 1cm
\begin{tabular}{ccccc} \hline
& $\sqrt{s}$ (GeV) &
$\sqrt{\frac{<Z^2>}{<N>}}$ (MeV) & $\sqrt{<z^2>}$ (MeV)  & $\Phi$ (MeV)
\\ \hline
p-p & 19.4 & 244.5 & 235.5 & 9.0 \\ \hline
p-Pb & 19.4 & 243.5 & 243.0 & 0.5 \\ \hline
Pb-Pb & 19.4 & 265.6 & 263.2 & 2.4 \\ \hline
\hline
p-p & 200 & 387.0 & 310.6 & 76.4 \\ \hline
p-Pb & 200 & 433.7 & 367.8 & 65.9 \\ \hline
Pb-Pb & 200 & 508.9 & 429.4 & 79.5 \\ \hline
\hline
p-p & 540 & 450.1 & 323.6 & 126.5 \\ \hline
p-Pb & 540 & 524.3 & 397.7 & 126.6 \\ \hline
Pb-Pb & 540 & 622.6 & 475.2 & 147.4 \\ \hline
\hline
p-p & 1000 & 455.5 & 324.4 & 131  \\ \hline
p-Pb & 1000 & 524.5 & 397.5 & 127 \\ \hline
Pb-Pb & 1000 & 704.2 & 484.3 & 220 \\ \hline
\hline 
\end{tabular}
\end{center}

\newpage

\newpage

\centerline{\bf Figure 1}
\vspace{1cm}

\begin{center}
\hspace{1.2cm}\epsfig{file=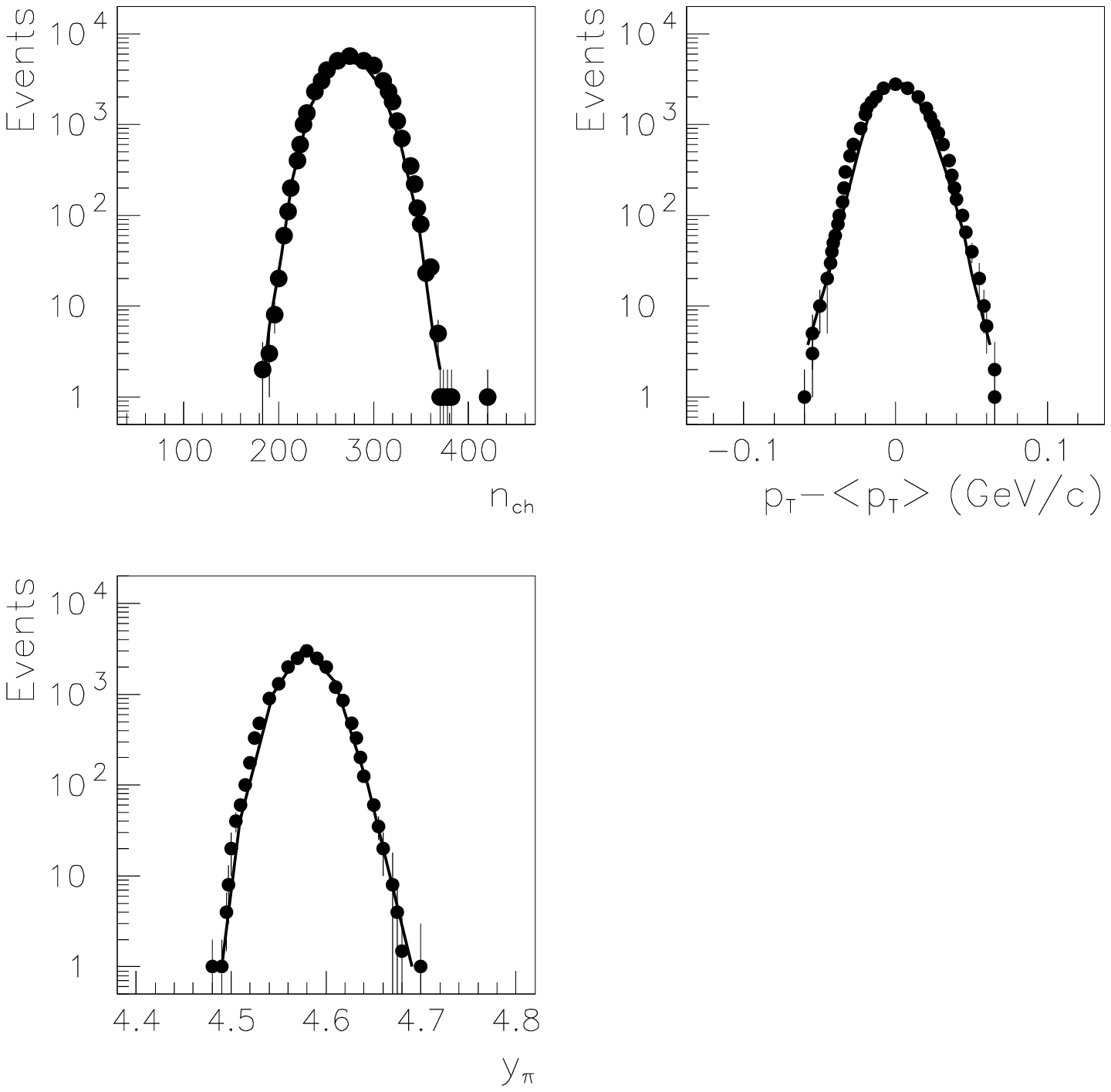,width=19.cm}
\end{center}

\newpage

\centerline{\bf Figure 2}
\vspace{1cm}

\begin{center}
\hspace{1.2cm}\epsfig{file=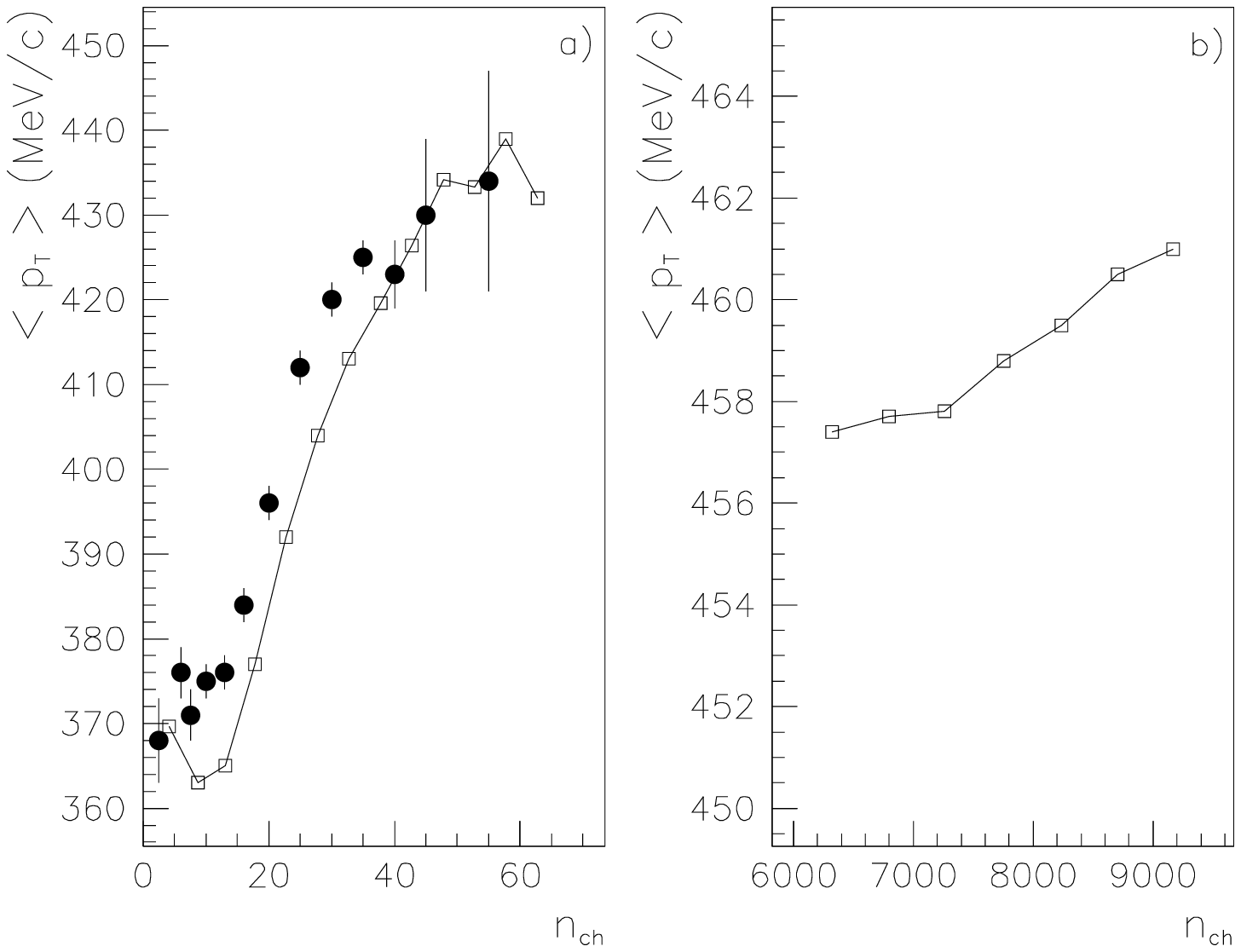,width=19.cm}
\end{center}

%\newpage
%
%\centerline{\bf Figure 3}
%\vspace{1cm}
%
%
%\begin{center}
%\hspace{-1.2cm}\epsfig{file=fig3.eps,width=14.cm}
%\end{center}

\newpage

\end{document}